\documentclass[11pt,twoside]{article} 
\usepackage{asp2004}
\usepackage{epsf}
\usepackage{psfig}
\usepackage{lscape} 

\begin{document} 
\title{Do Magnetic Fields Prevent Hydrogen from Accreting onto Cool Metal-line White Dwarf Stars?}  
\author{S. Friedrich} 
\affil{Max-Planck-Institut f\"ur Extraterrestrische Physik, 
    Giessenbachstr., 85748 Garching, Germany}
\author{S. Jordan} 
\affil{ Astronomisches Rechen-Institut, M\"onchhofstr. 12-14, 
    69120 Heidelberg, Germany}
\author{ and D. Koester} 
\affil{Institut f\"ur Theoretische Physik 
    und Astrophysik, Universit\"at Kiel, 24098 Kiel, Germany}
\begin{abstract} 
It is generally assumed that metals detected in the spectra of
a few cool white dwarfs cannot be of primordial origin and must
be accreted from the interstellar medium. However, the observed abundances
of hydrogen, which should also be accreted from the interstellar medium,
are lower than expected from metal accretion.
Magnetic fields are thought to be the reason for this discrepancy.
We have therefore obtained circular polarization spectra of the 
helium-rich white dwarfs GD\,40 and L745-46A, 
which both show strong metal lines as well as
hydrogen. Whereas L745-46A
might have a magnetic field of about $-$6900~G, which is about two times the
field strength of 3000~G necessary to repell hydrogen at the Alf\'en radius, 
only an upper limit for the field strength of GD\,40 of 4000~G (with 
99\% confidence) can be set which is far off the minimum field strength 
of 144000~G to repell hydrogen.
\end{abstract}

\section{Introduction}
A few helium-rich white dwarfs at the cool end of the white dwarf
sequence show evidence for metal lines in their spectra. Since radiation 
forces are not strong enough to compete with gravity for temperatures below
40000~K it is expected that metals sink down on time scales which are
short compared to the cooling age. If nevertheless metals are 
observed in the atmospheres, 
they must have come from outside the star. The most popular mechanism 
for this is accretion from interstellar matter.
The observed metal abundances in helium-rich 
white dwarfs are in agreement with accretion in solar 
element proportions (Dupuis et al. 1992, 1993a, 1993b). However, the 
observed abundances of hydrogen are much too low to be compatible 
with the accretion rates inferred from metal accretion. 
The mechanism most widely discussed as 
reason for this ``hydrogen screening'' is the propeller mechanism
adopted to white dwarfs by Wesemael \& Truran (1982): Metals are accreted 
in the form of grains onto a slowly rotating, weakly magnetized 
(10$^3$-10$^5$~G) white dwarf, whereas ionized hydrogen is 
repelled at the Alfv\'en radius.

\section{Observation}
Flux and circular polarization spectra of the white dwarfs GD\,40 
and L745-46A were observed with the VLT-UT1 and PMOS. 
Grism 600R (dispersion 1\AA /pix) with order sorting filter 
GG435 was used resulting in a spectral range of about 5200--7300\AA . 
The signal-to-noise 
ratio in the vicinity of H$\alpha$ and the helium lines of GD\,40 amounts to 
190 and 260, respectively, and to 280 for H$\alpha$ of L745-46A. The 
mean circular polarization in a spectral range between 5500\AA\ and 
6800\AA\ amounts to 0.02\%, and 0.03\% for GD\,40, and 
L745-46A, respectively, one order of magnitude lower than the expected 
values for kilogauss magnetic fields.

\section{Determination of Magnetic Fields}
\subsection{Weak-Field Approximation}
According to the theory of line formation in a weak magnetic field 
the splitting 
of a spectral line is proportional to the mean longitudinal
field $B_{\rm l}$. 
Provided that the Zeeman splitting of a spectral line is small
compared to intrinsic (thermal and pressure) broadening (e.g. Angel \&
Landstreet 1970) --- which is the case for white dwarfs and field strengths of
up to about 10kG -- the amount of polarization can be determined 
by the weak-field approximation, which is valid even in the presence of 
instrumental broadening, but is not generally correct in
the case of rotational broadening (Landstreet 1982):
$$\frac{V}{I}=- g_{\rm eff}\cdot 4.67\cdot 10^{-13} \lambda^2
\frac{dI}{I_\lambda d\lambda}B_{\rm l}$$
with $g_{\rm eff}$ the effective Land\'e factor which equals 1 for hydrogen
Balmer lines (e.g. Bagnulo et al. 2002).
For the HeI line at 5875\AA\ $g_{\rm eff}$ = 1.16 (Leone et al. 2000).

\subsection{Fitting Procedure}
In order to determine the mean longitudinal component of the magnetic field
the observed circular polarization was compared to the predictions of the
weak-field approximation in an interval of $\pm$20\AA\ around H$\alpha$ and
the HeI lines at 5875\AA\ and 6678\AA . The best-fit for $B_{\rm l}$, the only
free parameter, was found by a $\chi^2$ minimization procedure (see Aznar 
Cuadrado et al. 2004 for details).  

\subsection{Propeller Mechanism}
Minimum magnetic field strengths necessary for the propeller mechanism to work
were calculated according to Wesemael \& Truran (1982). 
Assuming a white dwarf mass of 0.6M$_{\odot}$, a radius of
0.013 R$_{\odot}$, an accretion rate of 10$^{-15}$ M$_{\odot}$/yr the 
corresponding minimum magnetic field strengths from their Eq.~4a are 
144000$^{+25000}_{-17000}$~G and 3000$\pm$500~G for GD\,40 and L745-46A, 
respectively. The error accounts for a reading error of 1mm in their Fig.~1. 
Further errors are introduced by the stellar radius which enters the formula 
with third power and the accretion rate. 

\section{GD\,40: Circular Polarization}
The inspection of the individual polarization spectra of GD\,40 did not show
any obvious features or variations in the polarization which cannot be 
attributed to noise. For the H$\alpha$ line the field
strength and the respective 68.3\%/99\% confidence ranges from the weak-field
approximation amount to 
(2747$\pm$6892/15000)G and for the two HeI lines at 
5875\AA\ and 6678\AA\ to ($-$1683$\pm$1528/3936)~G and 
($-$716$\pm$1848/4758)~G, 
respectively. A fit to all three lines
results in a magnetic field strength of 1131~G and a 99\% confidence level of
$\pm$ 2973~G. For comparison the derived field strength for a field star
amounts to $-$438$\pm$318/819~G (68.3\%/99\% confidence range). 
Thus we conclude that GD\,40 does not posses a magnetic field with an 
upper limit of about 4000 G. 

\section{L745-46A: Circular Polarization}

\begin{figure}
\plottwo{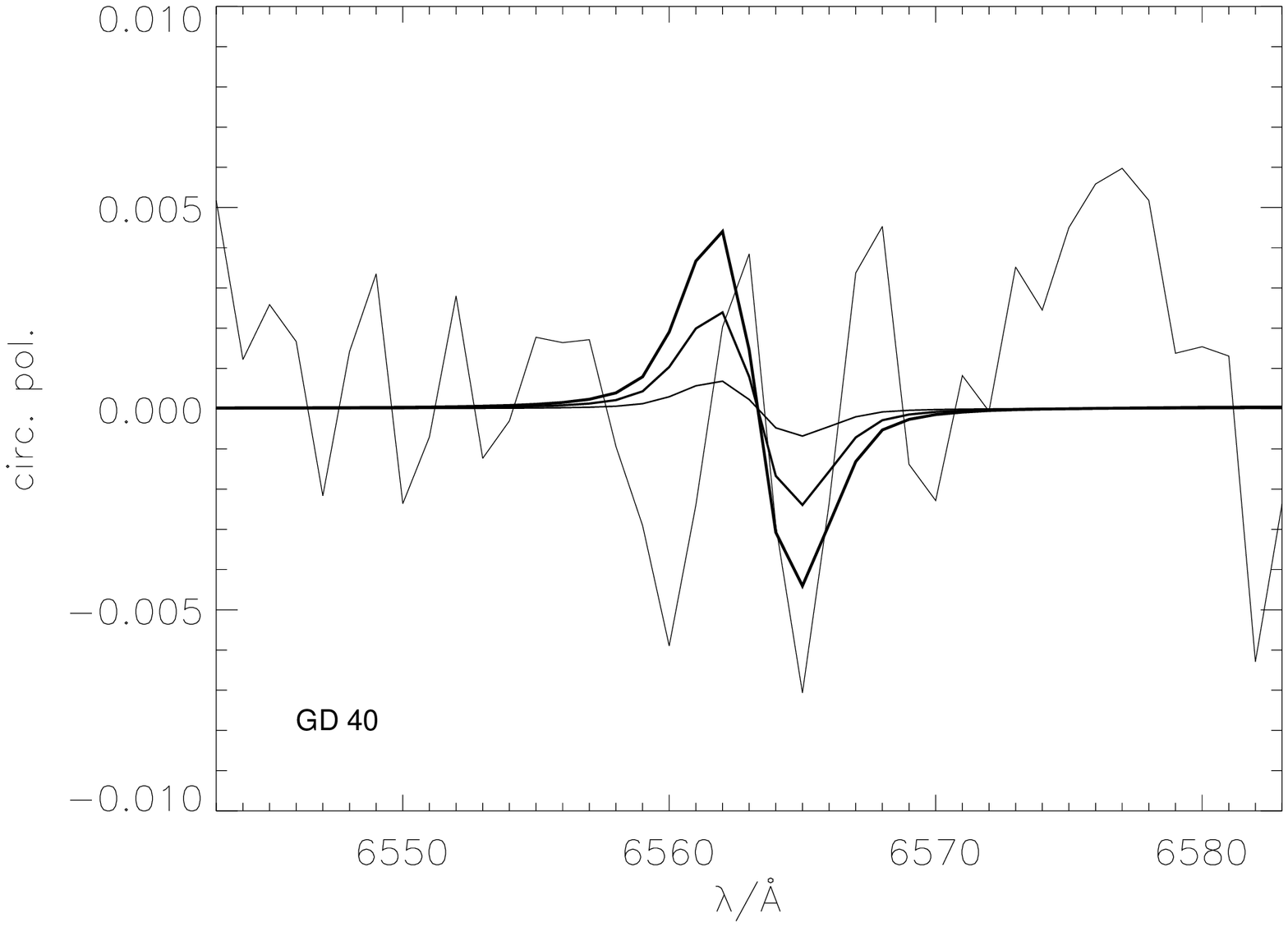}{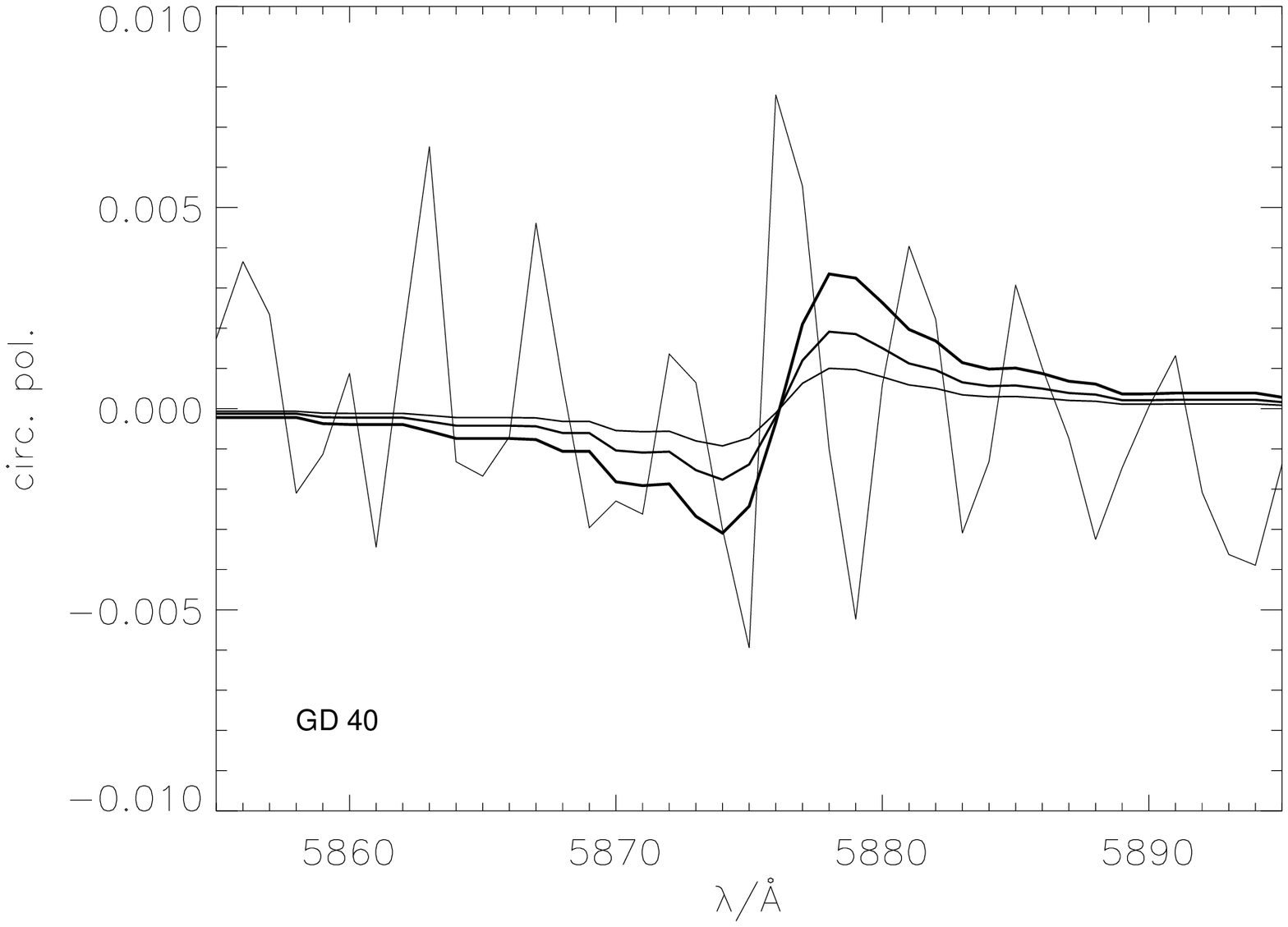}
\plottwo{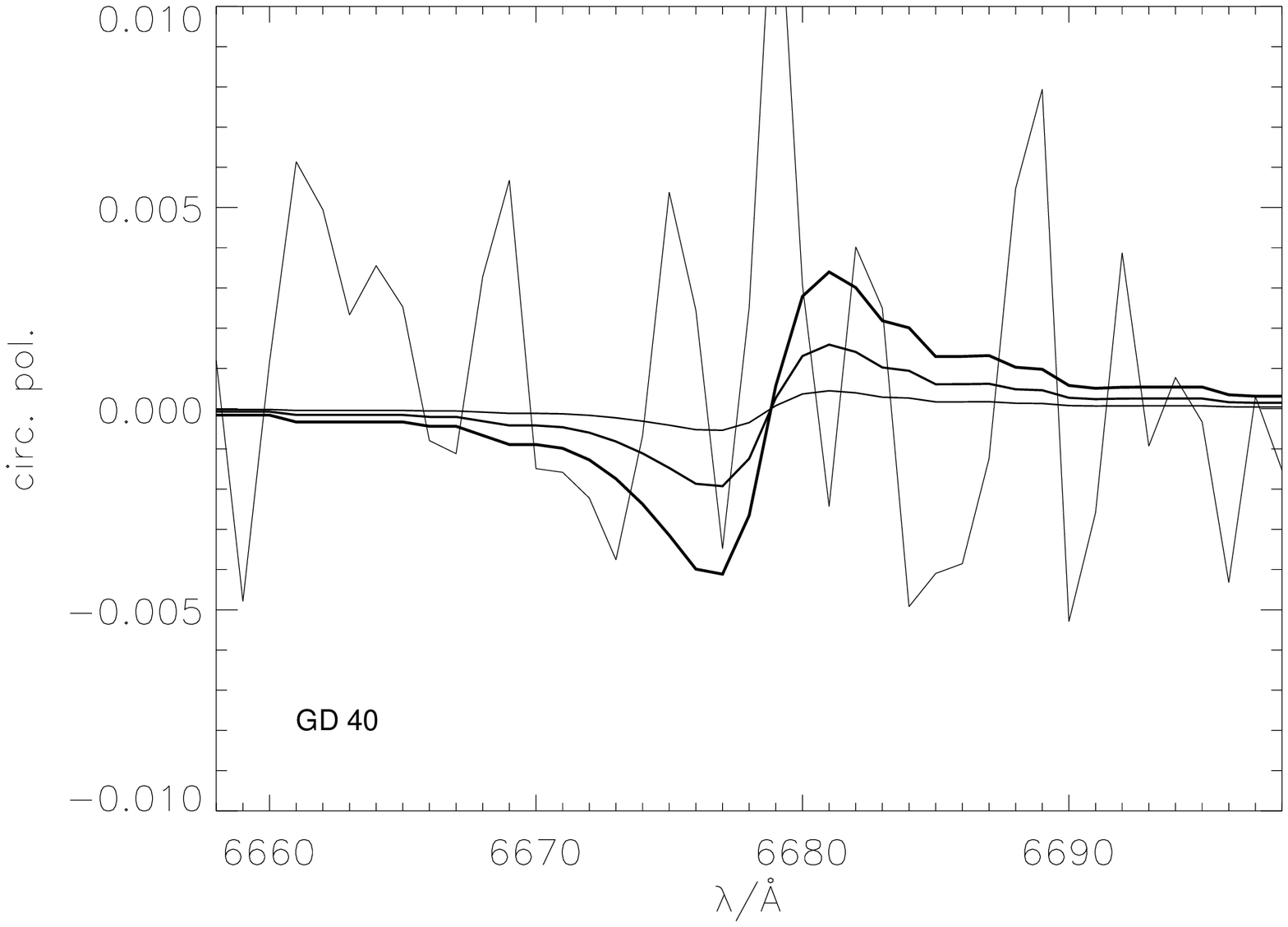}{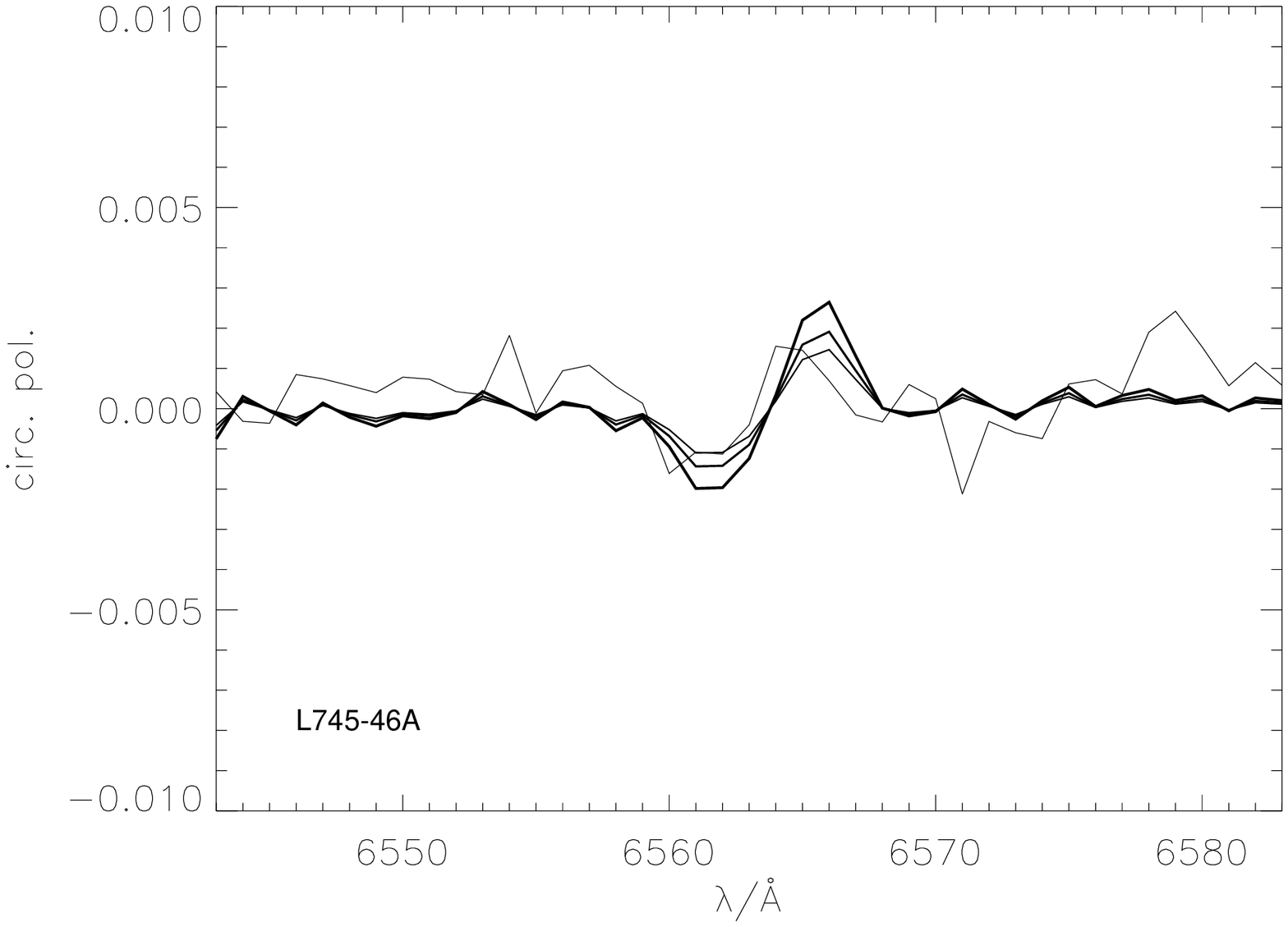}
\caption{Observed (gray) circular polarization spectra of GD\,40 and L745-46A 
together with the
  respective polarization spectra (bold) predicted by the weak-field
  approximation for field strengths of 2747~G, 9639~G, and 17747~G for 
  H$\alpha$, $-$1683~G, $-$3211~G, and $-$5619~G for the HeI line at 5875\AA ,
  $-$716~G, $-$2564~G, and $-$5474~G for the HeI line at 6678\AA\ (all GD40), 
  and $-$6900~G, $-$9000~G, and $-$12450~G for H$\alpha$ (L745-46A). 
  For each spectral line the first field
  strength denotes the best fit value, the second and third value the 
  best fit value plus the 68.3\%  and 99\% confidence range, respectively.}
\end{figure}

The best-fit magnetic field strength predicted by the weak-field 
approximation for the spectral range of H$\alpha$ is $-$6900~G with 
68.3\% and 99\% confidence ranges of $\pm$2100~G and $\pm$5550~G. 
Although the field strength clearly exceeds the statistical 
1$\sigma$ error one has to be cautious, because the result solely 
depends on the vicinity of the H$\alpha$ line. However, if this field could 
be confirmed, it would be strong enough to prevent hydrogen from accreting 
according to the prediction from the propeller mechanism of 3000~G.  

\section{Conclusion}
For L745-46A the predicted magnetic field strength from the weak field
approximation amounts to $-$6900~G with a formal 1$\sigma$ error of 
2100~G and a 99\% statistical confidence range of $\pm$5550~G. 
This magnetic field strength exceeds   
the minimum field strength of 3000~G necessary for the propeller 
mechanism to work. However, one should keep in mind, that this result is 
based on the H$\alpha$ line only. If the detection is confirmed, this means 
for the first time an indication that magnetic fields may play a role in
the accretion of hydrogen and metals onto cool helium-rich white dwarfs. 

However, we could not detect signatures of a magnetic field in the circular
polarization or flux spectrum of GD\,40. This does not necessarily 
mean, that 
no magnetic field is present, because if we are looking on the magnetic 
equator of a magnetic dipole ($i$=90$^\circ$), the components of the magnetic 
field along the line of sight completely cancel and no circular polarization 
can be detected. According to Wesemael \& Truran (1982) a field 
strength of about 144000~G is
needed for GD\,40 to let the propeller mechanism work. If we now assume that
we could detect a field strength of 4000~G (derived upper limit field 
strength from the fit to all three lines with a confidence of 99\%) we 
could estimate an inclination angle of greater than 84$^\circ$ for which 
the longitudinal field strength becomes to low to be detected (Brown et 
al. 1977). Therefore the chance to miss a magnetic field of 
144000~G is only about 6\%. From our observations we must therefore 
conclude that in GD\,40 probably other mechanisms than magnetic 
fields prevent hydrogen from accreting. More details can be found in 
Friedrich et al. (2004).

\acknowledgements{Work on white dwarfs in Kiel was supported by the DFG 
(KO-738/7-1); it is supported in T\"ubingen by the DLR (50 OR 0201).
}

\end{document}